\documentclass[aps,prl,tightenlines,twocolumn,superscriptaddress,showpacs]{revtex4}

\usepackage{amssymb}
\usepackage{color,graphicx}
\usepackage{amsmath}
\usepackage{amsbsy}
\usepackage{amsthm}
\usepackage{bbm}
\usepackage{bm}
\usepackage{epsfig}

\begin{document}

\title{Emergence of Molecular Chirality by Vibrational Raman Scattering}

\author{Farhad Taher Ghahramani}
\email[]{farhadtq@ch.sharif.edu}

\author{Afshin Shafiee}
\email[Corresponding Author:~]{shafiee@sharif.edu}
\affiliation{Research Group On Foundations of Quantum Theory and Information,
Department of Chemistry, Sharif University of Technology
P.O.Box 11365-9516, Tehran, Iran}

\begin{abstract}
  In this study, we apply the monitoring master equation describing decoherence of internal states to an optically active molecule prepared in a coherent superposition of non-degenerate internal states in interaction with thermal photons at low temperatures. We use vibrational Raman scattering theory up to the first chiral-sensitive contribution, i.e., the mixed electric-magnetic interaction, to obtain scattering amplitudes in terms of molecular polarizability tensors. The resulting density matrix is used to obtain elastic decoherence rates.
\end{abstract}


\maketitle

\emph{Introduction.}
Chirality is a fundamental concept in molecular physics and chemistry. Chiral molecules are stable, but not found in symmetric stationary states. In 1927, Hund explained stability of chiral states (and hence instability of their superposition) by a double-well potential model~\cite{Hun}. In Hund's model, chiral states are assumed to be localized in two minima of the potential. The superposition of chiral states is realized by tunneling between these two minima. However, Hund's approach seems unsatisfactory for some stable chiral molecules~\cite{Jan}. The problem can be addressed by introducing parity-violating terms in the molecular Hamiltonian~\cite{Rei,Heg,Let,Har,Qua,Wes} or non-linear terms due to the interaction with the environment, known as Decoherence program~\cite{Joo,Sim,Pfe,Fai,Sil,Has,Par,Ber,Hst1,Hst2,Hst3,Tro} (for a rather complete treatment see~\cite{sch}). The former, despite its small effect, can stabilize chiral states, if it would be larger than the inversion frequency, which is the case for many biologically stable chiral molecules, but the latter has received many attentions.

According to the decoherence theory, properties relating to molecular structure like chirality emerge after the interaction of the molecule with the environment~\cite{Sim,Hst1,Hst2,Hst3,Joo}. A molecule is generally described by translational and internal states, and environment is often modeled as a background gas or thermal photons. The theory of collisional decoherence for a particle with internal states is an extension of the positional decoherence of a particle without internal states~\cite{Gal,Dio,Dod,Hor1,Hor2,Hor3,Hor4,Vac1,Hor5,Vac2}. Hornberger derived a master equation describing internal quantum dynamics of an immobile system~\cite{Hor3} in the so-called monitoring approach (hereafter monitoring master equation), and Vacchini considered decoherence of translational and internal states of a system interacting with an inert gas~\cite{Vac1}. Trost and Hornberger applied the monitoring master equation to decoherence of chiral states of optically active molecules affected by a background gas~\cite{Tro}. Their basic idea is that an initially chiral molecule is blocked in that state through repeated scattering by a host gas.

Here, by using monitoring approach of collisional decoherence, we study the chiral stabilization of optically active molecules with internal states by thermal photons. The intermolecular effects are assumed to be negligible, which is the case for a dilute chiral media. Our discussion is limited to low temperatures at which two first states of contortional vibration (responsible for transforming between chiral configurations) are available. This is valid in most cases of interest. The initial state of the molecule is expressed by a coherent superposition of contortional states. Unlike Trost and Hornberger approach~\cite{Tro}, we assume that the initial superposition does not necessarily correspond to any chiral configuration. Then, we show that chirality of the molecule emerges due to the interaction with the beam of photons.

After a brief introduction to monitoring approach, we derive the monitoring master equation for an immobile two-state system (hereafter implicit master equation). The differential cross-sections appeared in the implicit master equation can be related to the vibrational Raman cross-sections. The theory of vibrational Raman scattering of optically active molecules was first presented by Atkins and Barron, based on polarizability tensors, as an extension of Kramers-Heisenberg formula~\cite{Atk}. We calculate the vibrational Raman cross-sections up to the discriminatory mixed electric-magnetic interaction. The resulting master equation (hereafter explicit master equation) is used to obtain elastic scattering rates.

\emph{Monitoring Master Equation.}
 Let us first explain the most widely used form of incorporating the environment i.e., the weak coupling approach. Long before and long after the collision, particles are well-separated and then evolution of the whole system is governed by the Hamiltonian $\hat {H}_{\mbox{\tiny$\circ$}}=\hat {H}^S_{\mbox{\tiny$\circ$}}+\hat {H}^E_{\mbox{\tiny$\circ$}}$ where $\hat {H}^S_{\mbox{\tiny${\mbox{\tiny$\circ$}}$}}$ and $\hat {H}^E_{\mbox{\tiny$\circ$}}$ are Hamiltonians of the system and the environment, respectively. Then, the total state at time $t$ after scattering is obtained by
\begin{equation}
\label{Eq:1}
|\psi(t)\rangle=\hat {U}|\psi_{sca}\rangle
=\hat {U}\hat {S}|\psi_{inc}\rangle
\end{equation}
where $\hat {U}=\exp(-{\frac{\imath \hat {H}_{\mbox{\tiny$\circ$}} t}{\hbar}})$ and $\hat{S}$ is the scattering operator. The $S$ matrix is characterized by the interaction Hamiltonian. In the weak coupling approach, interaction is weak, so that a perturbative treatment of the interaction is permissible. The monitoring approach, on the other hand, describes the environmental coupling non-perturbatively by picturing the environment as monitoring the system continuously, i.e., by sending probe particles which scatter off the system at random times. The temporal change of the system is obtained by multiplying the rate of collisions to the state transformation due to a single scattering. In this approach, the time evolution of the density matrix of the system is characterized by~\cite{Hor3}
\begin{align}
\label{Eq:2}
\partial_{t}\rho^{S} &=\frac{1}{\imath\hbar}[\hat {H}^S_{\mbox{\tiny$\circ$}},\rho^S]+Tr_{E}\Big(\hat {T}\hat\Gamma^\frac{1}{2}\rho^S\otimes\rho^E\hat\Gamma^\frac{1}{2}\hat {T}^{\dagger}\Big )\nonumber \\ & \quad +\frac{\imath}{2}Tr_{E}\Big[\hat\Gamma^\frac{1}{2}Re(\hat {T})\hat\Gamma^\frac{1}{2},\rho^S\otimes\rho^E\Big] \nonumber \\
& \quad -\frac{1}{2}Tr_{E}\Big \{\hat\Gamma^\frac{1}{2}\hat {T}^{\dagger}\hat {T}\hat\Gamma^\frac{1}{2},\rho^S\otimes\rho^E\Big \}
\end{align}
where $[\:]$ and $\{\}$ stand for commutation and anti-commutation relations, respectively. The operator $\hat T$ is the nontrivial part of the two-particle $\hat S$ operator, $\hat {S}=\hat {I}+\imath \hat {T}$ describing the effect of a single collision between environmental particle and system. The operator $\hat\Gamma$ specifies the rate of collisions. In the next section, we apply this master equation to the scattering of a beam of photons from chiral molecules.

\emph{Implicit Master Equation.}
A chiral molecule transforms between two chiral configurations by a long-amplitude vibration known as contortional vibration. To characterize this vibration, we employ a two-dimensional approach, which is valid for most molecular systems at low temperatures~\cite{Leg}. In this approach, chiral molecule is effectively described by a symmetric double-well potential with two minima. If we denote the small-amplitude vibration in each well by ${\omega}_{\mbox{\tiny$\circ$}}$, and the potential height by ${V}_{\mbox{\tiny$\circ$}}$, in the limit ${{V}_{\mbox{\tiny$\circ$}}}\gg\hbar{\omega}_{\mbox{\tiny$\circ$}}\gg k_{\mbox{\tiny$B$}}T$ (where $T$ is the temperature of the bath and $k_{\mbox{\tiny$B$}}$ is the Boltzmann constant), the first two states of the contortional vibration energy are available. The monitoring master equation is considerably made simple under this assumption. Here, we assume that the interaction of the chiral molecule with the beam of photons does not lead to any recoil of momentum of the molecule, only internal states are changed. This would be the case for massive molecules, in which translational degrees of freedom are fully decohered and therefore dynamics of them can be neglected. Then, the initial state of the molecule $\rho^{M}_{inc}$ for the relevant dynamics can be written as a superposition of first two states of contortional vibration
\begin{equation}
\label{Eq:3}
\rho^{M}_{inc}=\sum^{2}_{\nu,\nu'=1}c_{\nu}c^{\ast}_{\nu'}|\nu \rangle\langle \nu'|
\end{equation}
where $|\nu \rangle$ is the energy eigenstate, or a "channel" in the language of standard scattering theory. Note that chiral states are the maximal superposition of two corresponding channels.

The diagonal representation of the density matrix of the incident photons can be expressed as
\begin{equation}
\label{Eq:4}
\rho^{P}_{inc}=\frac {(2\pi\hbar)^{3}}{V}\int d\textbf{k} \mu(\textbf{k})|\eta (\textbf{k},n) \rangle\langle \eta (\textbf{k},n)|
\end{equation}
where $|\eta (\textbf{k},n) \rangle$ denotes the eigenstate of $\eta$ photons in the mode of momentum $\textbf{k}$ (normalized over box volume $V$) and circular polarization $n$. The momentum state of the incoming photons can be written as a phase space integration over projectors onto minimum uncertainty Gaussian wavepackets. Assuming black-body radiation, the momentum probability distribution of photons in unit volume could be expressed as
\begin{equation}
\label{Eq:5}
\mu(\textbf{k})d\textbf{k}=\frac{1}{4\pi^{3}\hbar^{3}N}\left(\frac{k^{2}}{e^{ck/k_{\mbox{\tiny$B$}}T}-1}\right)dkd\hat n
\end{equation}
with $N$ as the number of photons, $c$ as the speed of light and  $d\hat n$ as a solid angle differential in momentum space.

In the channel basis, time evolution of the reduced density matrix of the molecule is obtained as
\begin{equation}
\label{Eq:6}
\partial_{t}\rho^{M}=\sum_{\nu'',\nu'''}\partial_{t}\rho^{M}_{\nu''\nu'''}|\nu''\rangle\langle\nu'''|
\end{equation}
with matrix elements
\begin{align}
\label{Eq:7}
\partial_{t}\rho^{M}_{\nu''\nu'''} &=\Lambda_{\nu''\nu'''}\rho^M_{\nu'',\nu'''}+\sum_{\nu,\nu'}\rho^M_{\nu,\nu'}M^{\nu\nu'}_{\nu''\nu'''} \nonumber \\
&-\frac {1}{2}\bigg(\sum_{\nu}\rho^M_{\nu,\nu'''}\sum_{\nu^{(4)}}M^{\nu\nu''}_{\nu^{(4)}\nu^{(4)}}\nonumber \\ &+\sum_{\nu'}\rho^M_{\nu'',\nu'}\sum_{\nu^{(4)}}M^{\nu'''\nu'}_{\nu^{(4)}\nu^{(4)}}\bigg)
\end{align}
where
\begin{equation}
\label{Eq:8}
\Lambda_{\nu''\nu'''}=\frac {E_{\nu''}+\varepsilon_{\nu''}-(E_{\nu'''}+\varepsilon_{\nu'''})}{\imath\hbar}
\end{equation}
and $\varepsilon_{\nu''}$ is the energy shift of the molecule from energy $E_{\nu''}$ to $E_{\nu''}+\varepsilon_{\nu''}$. The rate coefficients are defined as
\begin{align}
\label{Eq:9}
M^{\nu\nu'}_{\nu''\nu'''}&=\int d\textbf{k}'\langle\eta(\textbf{k}',n')|\langle\nu''|\hat {T}\hat\Gamma^\frac{1}{2}|\nu\rangle\nonumber \\
& \qquad \qquad \qquad \rho^P_{inc}\langle\nu'|\hat\Gamma^\frac{1}{2}\hat {T}^{\dagger}|\nu'''\rangle|\eta(\textbf{k}',n')\rangle
\end{align}
The rate operator $\hat\Gamma$ is given by
\begin{equation}
\label{Eq:10}
\hat\Gamma=\sum_{\nu}|\nu\rangle\langle\nu|\otimes n_{\mbox{\tiny$P$}}c\sigma(k,\nu)
\end{equation}
where $n_{\mbox{\tiny$P$}}$ is the number density of photons, and $\sigma(k,\nu)$ is the total scattering cross section.

The elements of $T$-matrix are conveniently defined in terms of the multi-channel scattering amplitude $f$ as
\begin{align}
\label{Eq:11}
\hat {T}^{kk',nn'}_{\nu\nu'}=\frac{\imath c}{2\pi\hbar}\frac{f_{\nu\nu'}(\textbf{k},n;\textbf{k}',n')}{k}
 \delta(E_{\nu,k}-E_{\nu',k'})
\end{align}
where $E_{\nu,k}=ck+E_{\nu}$. At first sight, this leads to an ill-defined expression in terms of a squared delta function. However, conservation of the probability current implies a simple rule to deal with the squared matrix element ~\cite{Hor3,Hor4}. So, we have
\begin{align}
\label{Eq:12}
&\frac {(2\pi\hbar)^{3}}{V}\hat {T}^{kk',nn'}_{\nu\nu''}\hat {T}^{\dagger kk',nn'}_{\nu'\nu'''}\rightarrow c\chi^{\nu\nu'}_{\nu''\nu'''}\nonumber \\ & \frac{f_{\nu''\nu}(\textbf{k},n;\textbf{k}',n')f^{\ast}_{\nu'''\nu'}(\textbf{k},n;\textbf{k}',n')}{k^{2}\sqrt {\sigma(k,\nu)\sigma(k,\nu')}}\delta(E_{\nu'',k'}-E_{\nu,k})
\end{align}
with
\begin{equation}
\label{Eq:13}
\chi^{\nu\nu'}_{\nu''\nu'''} = \left\{
\begin{array}{rl}
1 & \text{if } E_{\nu''}-E_{\nu}=E_{\nu'''}-E_{\nu'} \\
0 & \text{otherwise}
\end{array} \right.
\end{equation}
Then, one obtains the rate coefficients as
\begin{align}
\label{Eq:14}
M^{\nu\nu'}_{\nu''\nu'''}&=n_{P}c^{2}\chi^{\nu\nu'}_{\nu''\nu'''}\int d\textbf{k}\mu({\textbf{k}})d\textbf{k}'\nonumber \\
& \qquad\frac{f_{\nu''\nu}(\textbf{k},n;\textbf{k}',n')f^{\ast}_{\nu'''\nu'}(\textbf{k},n;\textbf{k}',n')}{k^{2}}\nonumber \\ &\qquad \delta(E_{\nu'',k'}-E_{\nu,k})
\end{align}
Inserting rate coefficients into the density matrix of Eq.~(\ref{Eq:6}), after some mathematics we obtain
\begin{align}
\label{Eq:15}
\partial_{t}\rho^{M} &=\frac{n_{P}c^{2}}{2\pi^{3}\hbar^{3}}\int dkd\hat{n}dk'd\hat{n}'\frac{k'^{2}}{e^{ck/k_{\mbox{\tiny$B$}}T}-1}\nonumber \\ & \qquad \sum_{\nu\neq\nu'}\Big[(\rho^M_{\nu'\nu'}-\rho^M_{\nu\nu})|f_{\nu\nu'}|^{2}|\nu\rangle\langle\nu|\nonumber \\  & \qquad \quad -\rho^M_{\nu\nu'}\Big(|f_{\nu\nu}|^2+|f_{\nu\nu'}|^{2}\Big)|\nu\rangle\langle\nu'|\Big]
\end{align}
where $d\textbf{k}'=k'^{2}dk'd\hat {n}'$, and the right side is multiplied by $N$, the number of independent scattering events. Here, photon dependence of scattering amplitudes and corresponding energy conservations are implied for brevity.

In the case of elastic scattering, coherences are found to decay exponentially
\begin{equation}
\label{Eq:16}
\partial_{t}|\rho_{\nu\nu'}|=-\gamma^{ela}_{\nu\nu'}|\rho_{\nu\nu'}|
\end{equation}
The corresponding scattering rates are determined by the difference of scattering amplitudes
\begin{equation}\label{Eq:17}
\gamma^{ela}_{\nu\nu'}=\frac{n_{P}c^{2}}{4\pi^{3}\hbar^{3}}\int dkd\hat{n}dk'd\hat{n}'\frac{k'^{2}}{e^{\frac{ck}{k_{\mbox{\tiny$B$}}T}}-1}|f_{\nu\nu}-f_{\nu'\nu'}|^{2}
\end{equation}
In the next section, we calculate the corresponding scattering amplitudes.

\emph{Scattering amplitudes.}
The squared modulus of each scattering amplitude can be related to the corresponding Raman differential scattering cross-section as
\begin{equation}
\label{Eq:18}
|f_{\nu\nu'}(\textbf{k},n;\textbf{k}',n')|^2=4\pi^2\Big(\frac{k^2}{k'^2}\Big)\Big(\frac{d\sigma_{\nu\nu'}}{dn'}\Big)_{R}
\end{equation}
 Since molecule transforms between two chiral configurations by a vibration, the scattering amplitudes correspond to the vibrational Raman scattering, in which the interaction between chiral molecule and photon changes the vibrational state of the molecule (the electronic state of the molecule being unchanged) corresponding to the change of momentum and polarization of the photon. The corresponding contribution of the scattering amplitude is of the second order with two types of intermediate states, where there is absorption of one photon with momentum $\textbf{k}$ and circular polarization $n$, and emission of one photon with momentum $\textbf{k}'$ and polarization $n'$. Then, initial and final states can be written as $|\nu;\eta(\textbf{k},n)\rangle$ and $|\nu';(\eta -1)(\textbf{k},n),1(\textbf{k}',n')\rangle$. There are two types of scattering amplitudes, one-channel amplitudes ($f_{\nu\nu}$) and two-channel amplitudes ($f_{\nu\nu'}$). It is convenient to develop one-channel amplitudes of Rayleigh scattering first and then convert them to two-channel amplitudes of Raman scattering. In Rayleigh scattering, final state of the molecule is the same as initial state. The matrix element corresponding to the second order is obtained by
\begin{equation}
\label{Eq:19}
R_{\nu\nu}=\sum_{I}\frac{\langle f|I\rangle\langle I|\hat H_{int}|I\rangle\langle I|i\rangle}{E_{\nu}-E_{I}}
\end{equation}
where $|i\rangle$ and $|f\rangle$ are initial and final states, summation is over all possible intermediate states $I$, and $\hat H_{int}$ is the molecule-photon interaction Hamiltonian. The leading contribution of the scattering amplitude is purely electric in essence, occurring via electric dipole coupling, by which one cannot recognize optical activity. For chiral molecules, however, it is necessary to include the magnetic dipole coupling, leading to a relatively small chiral-sensitive mixed electric-magnetic contribution in the interaction Hamiltonian. The absolute value of the matrix element corresponding to the interaction of electric- ($\mu$) and magnetic-dipole moment ($m$) of the molecule with the corresponding fields of the light is obtained as~\cite{Cra}
\begin{align}
\label{Eq:20}
\big|R_{\nu\nu}\big|&=\Big(\frac{\hbar k}{2\varepsilon_{\mbox{\tiny$\circ$}}V}\Big)\eta^{\frac{1}{2}}\big|c {\hat n}'^{\ast}_{i}\hat n_{j}\alpha^{\nu\nu}_{ij}(k)\nonumber \\ & \qquad +{\hat n}'^{\ast}_{i}{(\hat k\times \hat n)}_{j}\beta^{\nu\nu}_{ij}(k)\mp \imath {\hat n}'^{\ast}_{i}\hat n_{j} {\beta}^{\nu\nu\ast}_{ji}(k)\big|
\end{align}
where unit vectors $\hat n$ and $\hat n'$ are incident and scattered polarization vectors and $\hat k$ is direction of momentum of the incident photon. Here, upper and lower minus/plus signs refer to left- and right-circular polarizations of the incident photon. Frequency-dependent electric polarizability and mixed electric-magnetic polarizability are defined as
\begin{align}
\label{Eq:21}
\alpha^{\nu\nu}_{ij}(k)=\sum_{r}\Big (\frac{\mu^{r\nu}_{i}\mu^{r\nu}_{j}}{E_{r\nu}-\hbar ck}+\frac{\mu^{\nu r}_{j}\mu^{\nu r}_{i}}{E_{r\nu}+\hbar ck}\Big)\nonumber \\
\beta^{\nu\nu}_{ij}(k)=\sum_{r}\Big (\frac{\mu^{r\nu}_{i}m^{r\nu}_{j}}{E_{r\nu}-\hbar ck}+\frac{m^{\nu r}_{j}\mu^{\nu r}_{i}}{E_{r\nu}+\hbar ck}\Big)
\end{align}
where $\mu^{r\nu}=\langle r|\mu|\nu\rangle$ and $m^{r\nu}=\langle r|m|\nu\rangle$, and $E_{r\nu}$ stands for the energy difference. Unlike the electric polarizability tensor $\boldsymbol{\alpha}$ (and its magnetic analogue), mixed electric-magnetic polarizability tensor $\boldsymbol{\beta}$ is parity-variant. Therefore, it can discriminate two chiral configurations.

The transition rates can be obtained by Fermi rule as
\begin{equation}
\label{Eq:22}
\Gamma_{\nu\nu}=\frac{2\pi}{\hbar}\rho|R_{\nu\nu}|^2
\end{equation}
where $\rho$ is the density of final states
\begin{equation}
\label{Eq:23}
\rho=\frac{Vk'^{2}dn'}{(2\pi)^{3}\hbar c}
\end{equation}
with $dn'$ as a solid angle around $k'$. Since in fluids, molecules are randomly oriented, transition rate is obtained by taking a rotational average~\cite{Cra}
\begin{equation}
\label{Eq:24}
\langle\Gamma_{\nu\nu}\rangle=\frac{4\pi\rho c\eta}{\hbar}\Big(\frac{\hbar k}{2\varepsilon_{\mbox{\tiny$\circ$}}V}\Big )^{2}A_{\nu\nu}
\end{equation}
with
\begin{align}
\label{Eq:25}
A_{\nu\nu} &=\mp Re\Big [\frac{\imath}{2}\Big(\delta_{jl}-\hat k_{j}\hat k_{l}\mp\imath\varepsilon_{jlm}\hat k_{m}\Big)\nonumber \\ &\qquad \qquad \qquad {\hat n}'^{\ast}_{i}{\hat n}'_{k}\Big(\langle\alpha^{\nu\nu}_{ij}\beta^{\ast\nu\nu}_{kl}\rangle+\langle\alpha^{\nu\nu}_{ij}\beta^{\nu\nu}_{lk}\rangle\Big)\Big]
\end{align}
where brackets denote the rotational average. Here, squared terms $\alpha^2$ and $\beta^2$ are vanished for random orientations, and elements of polarization of the incident photon was simplified as
\begin{equation}
\label{Eq:26}
\hat n_{i}\hat n^{\ast}_{j}=\frac{1}{2}\Big(\delta_{ij}-\hat k_{i}\hat k_{j}\mp\imath\varepsilon_{ijl}\hat k_{l}\Big)
\end{equation}
The rotational averaging of the fourth-rank tensor $\langle\alpha_{ij}\beta_{kl}\rangle$  is calculated by~\cite{Cra}
\begin{equation}
\label{Eq:27}
\langle\alpha_{ij}\beta_{i'j'}\rangle=I^{(4)}\alpha_{\mu\mu'}\beta_{\lambda\lambda'}
\end{equation}
with
\begin{equation}
\label{Eq:28}
I^{(4)}=\frac{1}{30}
\begin{bmatrix}
\delta_{ij}\delta_{i'j'}\\
\delta_{ii'}\delta_{jj'}\\
\delta_{ij'}\delta_{i'j}
\end{bmatrix}
^{T}
\begin{bmatrix}
4 & -1 & -1 \\
-1 & 4 & -1 \\
-1 & -1 & 4
\end{bmatrix}
\begin{bmatrix}
\delta_{\mu\mu'}\delta_{\lambda\lambda'}\\
\delta_{\mu\lambda}\delta_{\mu'\lambda'}\\
\delta_{\mu\lambda'}\delta_{\mu'\lambda}
\end{bmatrix}
\end{equation}
where $T$ means transpose, and Latin and Greek indices refer to space-fixed and molecular-fixed frames, respectively. Summation over repeated tensor suffices is implied. For the non-degenerate molecular states, polarizabilities $\boldsymbol\alpha$ and $\boldsymbol\beta$ can be chosen to be real and imaginary, respectively. So, inserting Eq.~(\ref{Eq:28}) into Eq.~(\ref{Eq:27}), and then into Eq.~(\ref{Eq:25}) one gets
\begin{align}
\label{Eq:29}
A_{\nu\nu} &=\pm\frac{1}{30}\Big[\big(|\hat n'.\hat k|^{2}\pm5|\hat k.\hat k'|-7\big)\alpha^{\nu\nu}_{\lambda\mu}\beta^{\nu\nu}_{\lambda\mu}+\nonumber \\ &\qquad \qquad \big(3|\hat n'.\hat k|^{2}\mp5|\hat k.\hat k'|+1\big)\alpha^{\nu\nu}_{\mu\mu}\beta^{\nu\nu}_{\lambda\lambda}\Big]
\end{align}
The differential cross-section is obtained by dividing the transition rate to the incident flux of the photons $\eta c/V$
\begin{equation}
\label{Eq:30}
\Big(\frac{d\sigma_{\nu\nu}}{dn'}\Big)_{R}=\frac{k^{2}k'^{2}}{8\pi^{2}\varepsilon_{\mbox{\tiny$\circ$}}^{2}c}A_{\nu\nu}
\end{equation}
To obtain two-channel cross-sections, we extend the results to the case of Raman scattering. As in conventional Raman experiments, we assume that frequency of the incident photon is not near-resonance, i.e.  $|E_{r\nu}-\hbar ck|\gg0$. Then, corresponding polarization tensors for the case of Raman scattering after factoring out rotational transitions is obtained by
\begin{equation}
\label{Eq:31}
\alpha^{\nu\nu'}_{ij}=\langle \nu|\alpha^{\nu\nu}_{ij}|\nu'\rangle, \qquad \beta^{\nu\nu'}_{ij}=\langle \nu|\beta^{\nu\nu}_{ij}|\nu'\rangle
\end{equation}
where $\alpha^{\nu\nu}_{ij}$ and $\beta^{\nu\nu}_{ij}$ are the usual Rayleigh polarizability tensors which depends on the normal coordinates of nuclei for the relevant vibration. Then, Raman scattering cross-section for optically active molecules is obtained by substituting usual Rayleigh tensors with corresponding Raman tensors.

\emph{ The Explicit Master Equation.}
Inserting the differential cross-sections in Eq.~(\ref{Eq:30}) into Eq.~(\ref{Eq:18}) and then into Eq.~(\ref{Eq:15}), dynamics of density matrix of the molecule is obtained as
\begin{align}
\label{Eq:32}
\partial_{t}\rho^{M} &=\frac{n_{P}c}{4\pi^{3}\hbar^{3}\varepsilon_{\mbox{\tiny$\circ$}}^{2}}\int dkd\hat{n}dk'd\hat{n}'\frac{k^{4}}{e^{ck/k_{\mbox{\tiny$B$}}T}-1}\nonumber \\ &\quad \sum_{\nu\neq\nu'}\Big[(\rho^M_{\nu'\nu'}-\rho^M_{\nu\nu})A_{\nu\nu'}|\nu\rangle\langle\nu|\nonumber \\  & \qquad -\rho^M_{\nu\nu'}\big(A_{\nu\nu}+A_{\nu\nu'}\big)|\nu\rangle\langle\nu'|\Big]
\end{align}
Here, the fourth power dependence on $k$ (Rayleigh’s law) is appeared. The polarization of the scattered photon can be written as the linear superposition of linear polarizations $\hat n'=1/\sqrt{2}\big(\hat n'^{\parallel}\pm\imath\hat n'^{\perp}\big)$ where $\hat n'^{\parallel}$ and $\hat n'^{\perp}$ are linearly polarized basis vectors. Then, if $\theta$ is the angle between $\hat k$ and $\hat k'$, we have
\begin{align}
\label{Eq:33}
A_{\nu\nu} &=\pm\frac{1}{30}\Big[\big(\frac{1}{\sqrt{2}}\sin^{2}\theta\pm5\cos \theta-7\big)\alpha^{\nu\nu}_{\lambda\mu}\beta^{\nu\nu}_{\lambda\mu}\nonumber \\ &\qquad \qquad +\big(\frac{3}{\sqrt{2}}\sin^{2}\theta\mp5\cos\theta+1\big)\alpha^{\nu\nu}_{\mu\mu}\beta^{\nu\nu}_{\lambda\lambda}\Big]
\end{align}
The squared of the amplitudes are isotropic, depending only on the magnitude of $\textbf{k}$ and the scattering angle $\theta$. Then, we can carry out the angular integrations by
\begin{equation}
\label{Eq:34}
\int d\hat {n}d\hat {n}'\rightarrow 8\pi^{2}\int d(\cos \theta)
\end{equation}
and the momentum integral can be computed using the definition of the Riemann $\zeta$-function for integer $n$
\begin{equation}
\label{Eq:35}
\zeta (n)=\frac{1}{(n-1)!}\int^{\infty}_{0}dx\frac{x^{n-1}}{e^{x}-1}
\end{equation}
So, we finally obtain
\begin{align}
\label{Eq:36}
\partial_{t}\rho^{M}&=\frac{8n_{p}k^{5}_{\mbox{\tiny$B$}}T^{5}}{5\pi\hbar^{3}c^{4}\varepsilon_{\mbox{\tiny$\circ$}}^{2}}
\sum_{\nu\neq\nu'}\Big[(\rho^M_{\nu'\nu'}-\rho^M_{\nu\nu})B_{\nu\nu'}|\nu\rangle\langle\nu|\nonumber \\  & \qquad \qquad \quad -\rho^M_{\nu\nu'}\big(B_{\nu\nu}+B_{\nu\nu'}\big)|\nu\rangle\langle\nu'|\Big]
\end{align}
with
\begin{equation}
\label{Eq:37}
B_{\nu\nu}=\mp\big[\frac{38}{3\sqrt{2}}\alpha^{\nu\nu}_{\lambda\mu}\beta^{\nu\nu}_{\lambda\mu}-\frac{6}{\sqrt{2}}\alpha^{\nu\nu}_{\mu\mu}\beta^{\nu\nu}_{\lambda\lambda}\big]
\end{equation}
Inserting corresponding amplitudes into Eq.~(\ref{Eq:17}), and assuming no phase difference between $f_{\nu\nu}$ and $f_{\nu'\nu'}$, the elastic decoherence rates are obtained as
\begin{equation}
\label{Eq:38}
\gamma^{ela}_{\nu\nu'}=\frac{4n_{p}k^{5}_{\mbox{\tiny$B$}}T^{5}}{5\pi\hbar^{3}c^{4}\varepsilon_{\mbox{\tiny$\circ$}}^{2}}
\Big|B^{\frac{1}{2}}_{\nu\nu}-B^{\frac{1}{2}}_{\nu'\nu'}\Big|^{2}
\end{equation}
In order to estimate an order of magnitude for the decoherence rate, one should calculate electric polarizability $\alpha$ and electric-magnetic polarizability $\beta$ tensors for the two-state contortional vibration mode in Rayleigh scattering from a chiral media. The quantum mechanical calculations of Rayleigh optical activity can be employed for our purpose. Results quoted in the literature are usually expressed in terms of mean $(\alpha\beta)^{\nu\nu}$ and anisotropic $(\gamma^{2})^{\nu\nu}$ invariant observables~\cite{Bar}
\begin{align}
\label{Eq:39}
  (\alpha\beta)^{\nu\nu}&=\frac{1}{9}\alpha^{\nu\nu}_{\lambda\lambda}\beta^{\nu\nu}_{\mu\mu}  \nonumber \\
   (\gamma^{2})^{\nu\nu}&=\frac{1}{2}\big(3\alpha^{\nu\nu}_{\lambda\mu}\beta^{\nu\nu}_{\lambda\mu}-\alpha^{\nu\nu}_{\lambda\lambda}\beta^{\nu\nu}_{\mu\mu}\big)
\end{align}
The calculations show that the mean invariant $(\alpha\beta)^{\nu\nu}$ is usually 1-3 orders of magnitude smaller than the anisotropic invariant $\gamma^{2}$~\cite{Zub}, and polarizability of molecules at their vibrational excited states goes smoothly to larger values~\cite{Mar}. Then, polarization-dependent term in decoherence rate would be at the order of $(\gamma^{2})^{00}$. The order of magnitude of a typical $(\gamma^{2})^{00}/c$ ($c$ is the speed of light) is about $10^{-83}C^{2}V^{-2}m^{4}$~\cite{Zub}. So, after making explicit the temperature dependence of number density of photons, the order of magnitude of decoherence rate could estimated as $10^{-95}(T/K)^{8}s^{-1}$. This shows clearly that the environmental photons cannot cause any suppression of interference between ground and excised states. At low temperatures, the maximal superpositions of first two relevant molecular states are chiral states. Then, according to einselection rule, decoherence of molecular states is equivalent to the stabilization of chiral states.

\emph{Conclusion.}
Chemists and some physicists are using chirality in the classical sense, i.e., by presupposing a molecule to be in one particular chiral state. However, according to decoherence theory, classical properties like chirality emerge out as a consequence of the interaction with the environment. Based on this approach, we have explored the collisional decoherence of a chiral molecule prepared in a coherent superposition of non-degenerate internal states in interaction with thermal photons. The temperature is assumed low, so that first two states of the relevant vibration of the molecule would be available. The reduced density matrix of the molecule are obtained in Eq.~(\ref{Eq:15}) as an extension of monitoring master equation. The appeared differential scattering amplitudes are calculated using vibrational Raman scattering theory to obtain the final master equation in Eq.~(\ref{Eq:36}). The corresponding elastic decoherence rates are calculated in Eq.~(\ref{Eq:38}). According to its estimated value, one can claim that the chirality of a molecule is an emergent property resulted due to the interaction of the molecule with a beam of photons.

\emph{Acknowledgement.}
We acknowledge the financial support of Iranian National Science Foundation (INSF) for this work. The authors thank Kamal Hajian and Amin Rezaei Akbarieh for instructive comments on some physical points.

\end{document}